\documentclass[aps,prl,twocolumn,superscriptaddress,showpacs,nobalancelastpage]{revtex4}
\usepackage{graphicx}
\usepackage{amsmath}

\def\Msun{{\rm M_\odot}}
\def\hMsun{h^{-1}{\rm M_\odot}}

\def\pcm{p_{cm}}\def\kms{{\rm km}\,{\rm s}^{-1}}

\begin{document}

\title{Dark Matter Halos with Cores from Hierarchical Structure Formation}  

\author{Louis E. Strigari}
\affiliation{Center for Cosmology, Department of Physics and Astronomy, University of California, Irvine, CA 92697,
 USA}

\author{Manoj Kaplinghat}
\affiliation{Center for Cosmology, Department of Physics and Astronomy, University of California, Irvine, CA 92697,
 USA}

\author{James S. Bullock}
\affiliation{Center for Cosmology, Department of Physics and Astronomy, University of California, Irvine, CA 92697,
 USA}

\date{\today}

\begin{abstract}
  We show that dark matter emerging from late 
  decays ($z  \lesssim  1000$) produces a linear power spectrum identical to that of Cold
  Dark Matter (CDM) on all observationally relavant scales ($\gtrsim
  0.1$ Mpc), and simultaneously generates observable constant-density
  cores  in small dark matter halos. We refer to this class of models as  
  {\it meta}-Cold Dark Matter ({\it m}CDM), because it is born with
  non-relativistic velocities from the decays of cold thermal relics. 
  The constant-density cores are a result of the low phase-space
  density of $m$CDM at birth. Warm dark matter cannot produce similar
  size phase-space limited cores without saturating  the Ly$\alpha$
  power spectrum bounds.  
  Dark matter dominated galaxy rotation curves and stellar velocity
  dispersion profiles may provide the  best means to discriminate
  between $m$CDM and CDM.    
  {\it m}CDM candidates are motivated by the particle spectrum of supersymmetric and extra dimensional 
  extensions to the standard model of particle physics. 
\end{abstract}


 \maketitle


\section{Introduction}

 In the standard cosmological model, the bulk of the matter in the
universe is in the form of Cold Dark Matter (CDM). Structure in CDM is
built up hierarchically, from initial density perturbations that are
adiabatic and nearly  scale-invariant  \cite{adiabatic}. 
CDM provides a compelling description  of the universe 
on large scales, but it has not conclusively passed all observational
tests on small scales. 
Two well-known  shortcomings are  that simulations 
predict  Milky Way-sized    galaxies   should have nearly an order of
magnitude more dark  
matter substructures than observed satellites \cite{missing}, and density
profiles of low-mass galaxies tend to be less cuspy than those seen in
simulations
\cite{Simon05,McGaugh05,Zackrisson06,KuziodeNaray06} and are often
well-fit by constant-density cores. While the former
can be alleviated both by  astrophysical \cite{astro} and cosmological 
\cite{Kamionkowski99,ZB03} methods,  the  latter, if ultimately true, 
may require a more drastic modification of the CDM paradigm \cite{drastic}.

Warm dark matter (WDM) provides an alternative that may
alleviate the problems on small scales \cite{WDM,HD}.  
Indeed, we {\em expect} constant density cores in the centers of WDM
halos because the initial phase-space maximum can never be exceeded
\cite{TG}.   An additional consequence of the large WDM particle
velocities is that 
its free-streaming length is large.
This suppresses the power
spectrum on small scales.  One problem with WDM as an alternative to
CDM, however, is that models with velocities large  enough to  
produce observationally-relevant cores
are in conflict with measurement of the Lyman-$\alpha$ forest power spectrum
\cite{Strigari:2006ue}.   
Indeed,  Lyman-$\alpha$  forest  analyses  suggest that the linear
density fluctuation spectrum  is nearly identical to the CDM
prediction on scales larger than $\sim 1$ Mpc
\cite{Narayanan00,Viel05,Abazajian05,Seljak06,Viel06}. 

It is thus interesting to consider if any modification to CDM can
reproduce the pattern of hierarchical structure formation, and  
also create large phase-space limited cores in the halos of low-mass
galaxies. In this paper  we investigate such a dark matter model,
which we refer to as  
 ``{\it meta}-Cold Dark Matter''  ($m$CDM). We define $m$CDM as
particles that emerge relatively late in 
cosmic  time  ($z \lesssim  1000$)    and are  born  non-relativistic 
from  the decays  of   cold particles. 
As we will show, the $m$CDM model provides both large phase-space cores
{\em  and}   CDM-like   power spectra    on  $\gtrsim  0.1$ Mpc scales,
consistent with the most stringent measurements of the 
Ly$\alpha$ forest power spectrum \cite{Seljak06,Viel06}.
Specifically, $m$CDM is born late enough that the free-streaming scale
is small, but the velocity dispersion is large enough to give rise to
a reduced phase-space density. Even though $m$CDM velocities are
non-relativistic, the redshifted velocities today are larger than the
corresponding velocities for WDM.  
Previous studies of similar models considered decays with shorter
lifetimes $\sim  M_{pl}^2/M_{weak}^3$ (of order   a month) to {\em  
  relativistic} daughter particles \cite{Kaplinghat05,Cembranos05}.
In  these models the relationship between the phase-space density and 
cut-off scale in the power spectrum is similar to that of WDM. 

Dark matter in the $m$CDM class arises in many extensions  to  the standard
model of particle physics. For  example  in  supersymmetric   models
we can  consider   a 
sneutrino decaying into a neutrino and gravitino with a lifetime $\tau
\simeq  3.6 \times    10^8$  s  $   (100   \, {\rm  GeV}/\Delta   m)^4
m_{\tilde{G}}/{\rm  TeV}$  \cite{Feng03,Ellis04}, where $\Delta m$ is
the mass difference between the sneutrino and gravitino, and
$m_{\tilde{G}}$ is the gravitino mass.  A  similar relation 
holds for Kaluza-Klein WIMPs   decaying to gravitons in theories  with
extra  dimensions. For example, we can consider a 10 TeV gravitino produced from
sneutrino decays with a lifetime $\tau = 5 \times  10^{12}$ s, where
the velocity imparted to the gravitino is $v/c \sim 10^{-3}$.  

\section{Decay Kinematics} We assume  that the decays  take the form
CDM $\rightarrow$  $m$CDM + 
$\ell$, where the decaying CDM is non-relativistic,  $m$CDM is the dark matter
today,  and $\ell$ is  a light neutral  particle. After the epoch of
decay, $a_d$,  
the velocity of the daughter particle will be $v \simeq (\pcm/{m})(a_d/a)$,
where $m$ is the $m$CDM  mass and $\pcm$ is the center of mass
momentum of the daughter particles.  
We define the average primordial phase-space density for the $m$CDM as
$Q_p \equiv \bar{\rho}/\bar{\sigma}^3$, where $\bar{\rho}$  is the
mean density and $\bar{\sigma}$ is the velocity dispersion. 
Integrating  over the phase-space distribution function (presented
below) we find \cite{Kaplinghat05} 
\begin{equation}
Q_p = 10^{-6} \alpha \left [10^3 p_{cm}/m_{cdm} \right ]^{-3} \left
[10^3 a_d \right]^{-3} \,, 
\label{eq:Q}
\end{equation}
where  $\alpha     =  1.0  (0.8)$  for     decays  in  the  radiation
 (matter)-dominated era.  Here, and for the rest of the
discussion, the units of $Q$ are ${\  \rm \Msun\,pc^{-3}(\kms)^{-3}}$. 
Note that for the typical cases we consider, $(m_{cdm} - m)/m_{cdm} \simeq \pcm /m_{cdm} \simeq 10^{-3}$. 

In the case of late decays, the power spectrum is governed not only by
$Q_p$ but also by the epoch of decay, $a_d$.
We can understand this heuristically in terms of standard free-streaming
arguments.  
 In this approximation, power on a comoving scale smaller 
than a streaming scale,
\begin{equation}
\lambda_{FS} \simeq \int_{\tau}^{t_0} d t v(a)/a 
\equiv (p_{cm} a_d / m) \int_{a_d}^{1} d a a^{-3}/H(a)\,,
\end{equation}
will be erased.
Here $\tau  = t(a_d)$ is the decay  lifetime, $t_0$ 
is the current epoch, and we have assumed that the decay  is
non-relativistic.  If
the decay  were to occur well before  matter-radiation equality,
$a_d <   a_{eq}$,    we would have  $\lambda_{FS}   \simeq p_{cm} a_d /m
\ln(a_{eq}/a_d) a_{eq}^{1/2}  0.5   H_0^{-1}$, and  the filtering  
would be driven  primarily  by the phase-space density variable
$p_{cm} a_d/m$,  and only logarithmically on the decay lifetime through
 $a_d$.  If we consider decays in the matter-dominated regime, then the
power spectrum filtering scale depends on 
{\em both} the   decay  epoch and the   phase-space  density:
$\lambda_{FS} \simeq  (p_{cm} a_d/m) a_d^{-1/2}   0.5   H_0^{-1}$  ($
\simeq   0.05$ h$^{-1}$Mpc).  Thus for long  lifetimes  ($a_d \gtrsim
a_{eq}$),   the free-streaming scale is suppressed even if $Q_p$  
is relatively small. For the models we consider, the universe does not
re-enter a radiation dominated phase due to the decay.

\begin{figure}[t!]
\includegraphics[width=3.in,clip=true]{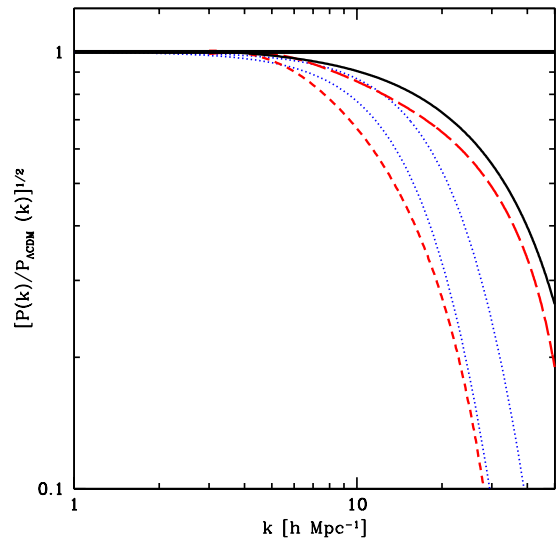}
\caption{ The  ratio of the power spectrum to that of standard
  $\Lambda$CDM  for a variety of different models. 
The red curves show $\Lambda+$$m$CDM models with $\tau = 5 \times
  10^{12}$ s and $Q_p = 10^{-5} $ 
(long-dash) and $Q_p = 10^{-6}$ (short-dash). The solid black curve is
  a $\Lambda+$$m$CDM model with  
$\tau = 10^{14}$ s and $Q_p = 10^{-6} $. Also shown are limits on the
  warm dark matter power spectrum   
from the analyses of \cite{Seljak06} (dotted, right) and
  \cite{Viel06} (dotted, left).  
\label{fig:tf}}
\end{figure}

\section{Power Spectrum} 

In order to track the evolution of fluctuations we will need the $m$CDM phase-space
distribution.  
Following \cite{Kawasaki92,Kaplinghat05}, we find that
\begin{equation}
f(a,q) = \frac{2 \pi^2 \rho(a) \exp(-t_q/\tau)}{q^2 p_{cm} \tau m}\left( \frac{1}{a_q H(a_q)}\right).  
\label{eq:fdm}
\end{equation}
The variables  $a_q  \equiv q/\pcm$, and $t_{q}  \equiv t(a_q)$ are
characteristic epochs and 
times for each value of the comoving momentum $q$. 
In the  limit of small $p_{cm}/m$,  we can write all of the
perturbations to the phase-space distribution in terms  of  the density 
and velocity perturbations. For  the decaying CDM,   the evolution  of these
quantities    is  identical to   the standard non-decaying case
\cite{Ma95}. This results from the fact that the  decays act to
remove  the same number  of particles from every region of  space. 

To compute the power spectrum we use the full hierarchy of equations
governing the perturbation of the phase-space 
distribution. We use the concordance cosmology \cite{Spergel06}, and
assume that the universe is  
made flat by a cosmological constant ($\Lambda$), and that the dark
matter is in the form of $m$CDM.  
We can gain insight into the effect  of the decays by applying energy
conservation and considering a simpler set of equations in the limits
where the $m$CDM equation of state and sound speed is negligible.
For  large scales, $k  \rightarrow   0$, we have $y^\prime  \equiv 
\delta^{\prime    \prime}         -\delta_{cdm}^{\prime       \prime}=
-y/\tau(1 - 2m_0^2\rho_{cdm}/m_{cdm}^2\rho)$, where $\delta$ is the
$m$CDM density perturbation and  
the derivatives here are with respect to cosmic time, related to the
conformal time, $\eta$, by $d \eta = dt/a$.  
We have defined $m_0^2 = m_{cdm}^2 - m^2$.
For small $a$, $y$ is driven to zero as $\eta^2$ and is exponentially
damped for $a > a_d$.  Thus the $m$CDM perturbation   `catches   up'
with  the  CDM 
perturbation  almost immediately,  and the  power spectrum  retains
the form of  standard $\Lambda$CDM on scales larger
than the horizon at decay. 

We expect departures from a $\Lambda$CDM power spectrum approximately when $k^2
c_s^2 \eta^2 \sim 1$, where $c_s^2$ is the $m$CDM sound speed. 
For  decays in the radiation-dominated regime, this damping scale is given by 
$p_{cm} a_d/m$, and  for  decays  in  the matter-dominated  regime  the scaling is 
$(p_{cm} a_d /m) \tau^{-1/3}$.    
Thus  the damping scale for matter-dominated decays is set by $Q_p$
and $\tau$, while 
for decays during the radiation era the scale depends only on $Q_p$. 
As one would expect, these are  the same scalings we obtained earlier
with our simple free-streaming analysis.  

In Figure \ref{fig:tf} we  show the ratio of the power spectrum for a
variety of $\Lambda+$$m$CDM 
models to $\Lambda$CDM. The dotted lines are warm dark
matter models that correspond to the bounds from recent analyses of
the Lyman-$\alpha$ forest  
\cite{Seljak06,Viel06}.  The model with $\tau =  5 \times 10^{12}$ s
and $Q_p = 10^{-6}$ 
is potentially inconsistent, though these bounds can clearly be evaded
with a $Q_p = 10^{-5}$ model.  
An example model with a longer lifetime of $10^{14}$ s and $Q_p = 10^{-6}$, 
however, is clearly consistent with both bounds, which is what we
predict from the above scalings. Further detailed 
modeling of the Lyman-$\alpha$ power spectrum on small scales thus
provides an excellent probe of viable $m$CDM models.  

\section{Phase-Space Cores} 

We can compare the primordial phase-space
values $Q_p$ to the phase-space densities deduced  
from measurements of galaxy rotation curves. 
Among the highest-resolution rotation curves  for low-mass spirals are
those  presented  by  \cite{Simon05} who use two-dimensional
H$\alpha$  and CO velocity fields.  
When the dark matter halo components  of these galaxies are
fit to cored density distributions, the average central phase-space densities
are $2 \times 10^{-6}$ \cite{Simon05,Strigari:2006ue}. $Q_p$ cannot be
smaller than this value.

In order to connect $Q_p$ to the maximum phase space density in a
halo, $Q_0$, we  must consider halo formation. 
We apply a general argument for collisionless collapse that is much
stronger than the Tremaine-Gunn bound for distributions with maximum
fine-grained phase space density much larger $Q_p$.
Specifically, the Excess Mass Function (EMF),   defined in
\cite{Dehnen05}, must  always decrease. Following the approach  of
\cite{Kaplinghat05}, we calculate the EMF of primordial $m$CDM using
Eq.  \ref{eq:fdm} and demand that its value for all phase-space
densities be greater than the EMF of the dark matter
halo profile. This procedure 
gives us the maximum central phase space density of the dark matter
halo profile that is consistent with a given primordial momentum 
distribution.

The procedure depends on the details of the decay process as well as
the assumed dark matter halo profile. We will restrict our attention
here to the Milky Way satellite galaxies, and this motivates
a general parameterization for the dark matter
halo \cite{Kazantzidis:2005su} that interpolates between a truncated NFW profile and a profile
with a core -- $\rho(r) \sim \exp(-r/b)/(c+r)/(r+a)^2$. For parameters
that are broadly consistent with the Milky Way
dwarf-satellite galaxies we find numerically that $Q_0 \lesssim 30 Q_p$
for $Q_p=10^{-5}$ and $Q_0 \lesssim 50 Q_p$ for $Q_p=10^{-6}$. 

The above procedure also provides an estimate of the
{\em minimum} core size. We find that the minimum value of $c$ ranges
between 300 to 800 pc for $Q_p=10^{-6}$ depending on the value of $a$
(restricted to be greater than $c$) for a halo with mass within 1 kpc
of $10^7 \Msun$ (which results in a central stellar velocity
dispersion $\sim$ 10 km/s for a typical satellite dwarf
galaxy). The lifetime is consistent with the model in Figure
\ref{fig:tf}. A decrease in halo mass to  $10^6 \Msun$ increases
minimum $c$ range to 1.5--2.5 kpc, while increasing mass to $10^8
\Msun$ decreases it to 150--300 pc range. The minimum $c$
scales approximately as $1/\sqrt{Q_p}$ in this range of parameter
space. 

We caution that the above exercise only provides us with the minimum
possible core size. The merging process is expected to increase this
core size \cite{Kazantzidis:2005su}
 and detailed simulations are required to make further
predictions. In the above example, we have allowed the inner core
($c$) and the overall scale radius of the halo ($a$) to vary freely of
each other. However, mergers
will certainly correlate the two scales. Also, since
halos get built out of smaller halos, the excess mass function
analysis should be applied to the smaller building blocks before
merger.

Hierarchically-formed dark  matter  halos from CDM particles without  
an appreciable phase-space maximum are known  to have power-law
phase-space   profiles,     $Q(r)     =   \rho(r)/\sigma^3(r)      =
Q_{-2}(r/r_{-2})^{-\alpha}$, with  $\alpha \simeq 1.875$ \cite{Taylor01}.
Here we have normalized the $Q$ profile at  the radius where the log-slope
in the density profile is $-2$.  We can use this result also to
estimate the size of the core radius, $r_{\rm core}$, that would be
imposed for a given central phase-space 
density.  A minimum estimate for the core size
will assume that the power-law continues uninterrupted
until the central $Q$ is reached,
 $Q(r=r_{\rm core}) = Q_0$.  
We must first determine the normalization constant $Q_{-2}$.
We do so assuming an NFW profile \cite{nfw96}
and find
$Q(r) \simeq 10^{-11} M_{11}^{-1} (c_{vir}/15)^{-0.125}(r/R_{\rm v})^{-1.875}$,
where $M_{11} \equiv (M_{\rm v}/10^{11} \hMsun)$ characterizes the
halo virial mass, 
$R_{\rm v} \simeq 134$~kpc~$M_{11}^{1/3}$  
is the halo virial radius, and
$c_{vir} \equiv R_{\rm v}/r_{-2} \simeq 15$ is appropriate for 
low-mass spiral galaxy halos \cite{Bullock01}.   The quoted power-law
dependence is 
approximate but serves to highlight that the dependence is expected to be weak.
If we use $Q(r_{\rm core}) = Q_{0}$ to derive a lower limit on the core size
and ignore weak $c_{vir}$ dependence we obtain
$r_{\rm core} \gtrsim 2.4 \times 10^{-3} R_{\rm v} M_{11}^{-0.53}
Q_{-6}^{-0.53}$,  
where  $Q_{-6}   \equiv   Q_{0}/(10^{-6})$
characterizes the $Q_{0}$  dependence.  Small spiral
galaxy halos   ($M_{11} \simeq  1$) with  $Q_{0}$ set    by the
Lyman-$\alpha$ forest limit for WDM ($Q_{-6} \simeq  8 \times 10^3$)
can have only small cores $r_{\rm core} \gtrsim  1$ pc.  A $m$CDM
candidate can avoid 
the Lyman-$\alpha$  forest bound  with   $Q_{-6} \gtrsim  10$  and produce
larger, dynamically-important cores, $r_{\rm core}  \gtrsim 100$ pc.
As we have emphasized, this estimate represents a conservative lower limit
on the full extent of the core region because we have assumed a sharp
break in the phase-space profile.  The WDM N-body simulation results
of  \cite{Avila00} suggest that the cores are $\sim 3$ times larger
than our estimate would give.   

Note also that the core size is expected to take up a larger fraction of
the halo as we consider smaller systems.  
Large, soft   cores will render these dwarf-size halos quite prone  to
disruption upon accretion into larger halos \cite{ZB03}.  In $m$CDM we
thus expect the predicted substructure count to be reduced relative to
CDM  predictions.   

\section{Discussion} 
For the models considered, the net effect of the injected relativistic
energy during decay may be phrased in terms of the effective number of
light neutrinos, $\Delta N_\nu =  1.8  \times    10^{-2}
\sqrt{\tau/{\rm yrs}} (p_{cm}/m)$. For the $m_{\tilde G} = 10$ TeV,
$\tau = 5 \times 10^{12}$ s model described above, we have $\Delta
N_{\nu}  \simeq 0.01$ which would be hard to detect.

Neutrinos  and    photons  produced  along   with  $m$CDM  will arrive
unscattered in the diffuse radiation backgrounds today.  The strongest
constraint  comes from  the  neutrinos produced  directly in  two-body
decays. The relevant low-energy  neutrino flux limit is   $< 1.2$
cm$^{-2}$  s$^{-1}$ from  Super-Kamiokande for energies $18-82$ MeV
\cite{SK}. The  above sneutrino model is 
consistent  with this limit and in the sensitivity range of future
experiments with  reduced    energy
thresholds \cite{gadzooks}. Photons may also be produced in three-body
decays and 
subsequent hadronization into neutral pions. Using a very conservative
branching  fraction of 10$^{-3}$   into photons \cite{Feng04}, we find
that the photon flux is consistent with observations \cite{Kinzer97}.

In  conclusion, we have  shown  if dark  matter  is produced from the late
decay of cold relics, it will give rise to large  cores in 
low-mass  galaxies {\em and}  alleviate  the dwarf satellite
problem,   without  destroying  agreement with   the observed
Lyman-$\alpha$ forest power spectrum.  This  differs from  standard WDM
models, which,  in   order  to remain consistent   with   constraints
from the Lyman-$\alpha$ power  spectrum,  cannot produce sizeable cores
in small galaxies. Dynamical studies of nearby galaxies 
may provide the best
means to test the $m$CDM scenario and compare it to CDM predictions. 


We  thank J. Beacom, J. Cooke, S. Kazantzidis, S.
Koushiappas, and R. Wechsler for  discussions. LES is supported in
part by a Gary McCue Postdoctroral Fellowship through the  Center for
Cosmology at UC Irvine.  We acknowledge the use of CMBfast \cite{cmbfast}.
%


\end{document}